\documentclass[pra,twocolumn,graphicx]{revtex4}
\usepackage{amssymb}
\usepackage{epsfig,amsmath}

\begin{document}

\title{One and two-center processes in high-order harmonic generation in
diatomic molecules: influence of the internuclear separation}
\author{C. Figueira de Morisson Faria}
\affiliation{Department of Physics and Astronomy, University College London, Gower
Street, London WC1E 6BT, United Kingdom}
\date{\today}

\begin{abstract}
We analyze the influence of different recombination scenarios, involving one
or two centers, on high-order harmonic generation (HHG) in diatomic
molecules, for different values of the internuclear separation. We work
within the strong-field approximation, and employ modified saddle-point
equations, in which the structure of the molecule is incorporated. We find
that the two-center interference patterns, attributed to high-order harmonic
emission at spatially separated centers, are formed by the quantum
interference of the orbits starting at a center $C_{j}$ and finishing at a
different center $C_{\nu }$ in the molecule with those starting and ending
at a same center $C_{j}.$ Within our framework, we also show that
contributions starting at different centers exhibit different orders of
magnitude, due to the influence of additional potential-energy shifts. This
holds even for small internuclear distances. Similar results can also be
obtained by considering single-atom saddle-point equations and an adequate
choice of molecular prefactors.
\end{abstract}

\maketitle

\section{Introduction}

Molecules in strong laser fields have attracted a great deal of attention in
the past few years. Indeed, strong-field phenomena, such as high-order
harmonic generation, above-threshold ionization, and nonsequential double
ionization, may be used as tools for measuring and even controlling dynamic
processes in such systems with attosecond precision \cite{Scrinzi2006}. This
is a direct consequence of the fact that the physical mechanisms behind such
phenomena take place within a fraction of the period of the laser field. For
a typical, titanium-sapphire laser used in experiments, whose period is of
the order $\tau \sim 2.7\mathrm{fs}$, this means hundreds of attoseconds.

Explicitly, such phenomena can be described as the laser-assisted
rescattering or recombination of an electron with its parent ion, or
molecule \cite{tstep}. At an instant $t^{\prime },$ this electron reaches
the continuum through tunneling or multiphoton ionization. Subsequently, it
propagates in the continuum, being accelerated by the external field.
Finally, it is driven back towards its parent ion, or molecule, with which
it recombines or rescatters at a later instant $t$. In the former case, the
electron kinetic energy is converted in a high-energy, XUV photon, and
high-order harmonic generation (HHG) takes place \cite{hhgsfa}. In the
latter case, one may distinguish two specific scenarios: The electron may
suffer an elastic collision, which will lead to high-order above-threshold
ionization (ATI) \cite{atisfa}, or transfer part of its kinetic energy to
the core, and release other electrons. Hence, laser-induced nonsequential
double (NSDI), or multiple ionization (NSMI) will occur.

For molecules, there exist at least two centers with which the electron may
recombine or rescatter. This leads to interference patterns which are due to
photoelectron or high-harmonic emission at spatially separated centers, and
which contain information about its specific structure. In the simplest case
of diatomic molecules, such patterns have been described as the microscopic
counterpart of a double-slit experiment \cite{doubleslit,KB2005}.

A legitimate question is what sets of electron orbits are most relevant for
the two- or many-center interference patterns. To understand this issue is a
first step towards controlling such processes by, for instance, an adequate
choice of the shape and polarization of the external field. In the specific
case of diatomic molecules, the electron may start and return to the same
center $C_{j}$, or leave from a center $C_{j}$ and return to a center $%
C_{\nu },\nu \neq j(j=1,2)$. Hence, in total, there exist four possible
processes that contribute to the yield. Recently, these processes have been
addressed in several studies, for above-threshold ionization \cite%
{Usach2006,HBF2007,DM2006,BCCM2007,Milos2008}, high-order harmonic
generation \cite{KBK98,KB2005,PRACL2006,F2007} and nonsequential double
ionization \cite{F2008}. The vast majority of these studies has been
performed using semi-analytical methods, in the context of the strong-field
approximation. In this framework, the transition amplitude can be written as
a multiple integral with a slowly varying prefactor and a semiclassical
action. The structure of the molecule may be either incorporated in the
former \cite{MBBF00,Madsen,Usachenko,Kansas,JMOCL2006,FSLY2008}, or in the
latter \cite{Usach2006,HBF2007,Milos2008,KBK98,PRACL2006,F2007,F2008}. On a
more specific level, when solving these integrals employing saddle-point
methods, it is possible to draw a space-time picture of the laser-assisted
rescattering or recombination process in question, and establish a direct
connection to the orbits of a classical electron in a strong laser field
\cite{orbitshhg}. By incorporating the structure of the molecule in the
action, one obtains modified saddle-point equations which gives the one-or
two-center scenarios.

In a previous publication \cite{F2007} , we have addressed this issue to a
large extent for high-order harmonic generation, within the Strong-Field
Approximation (SFA). \ Our results suggested that the maxima and minima
observed in the spectra were due to the quantum interference of the
processes in which the electron leaves and returns to a specific center $%
C_{j}$ in the molecule with those in which it leaves from $C_{j}$,
but returns to a different center $C_{\nu }.$ There exist, however,
a few ambiguities as far as the interpretation of our findings is
concerned. For instance, in the length-gauge formulation of the SFA,
we  found additional potential energy shifts, which depend on the
field strength $E(t)$ and in the internuclear separation $R.$ These
shifts led to a strong suppression of tunnel ionization at one of
the centers. This could have led to the conclusion that the
interference between other processes were not relevant for the
patterns in the spectra.

In this proceeding, we investigate the role of the one and two-center
recombination scenarios in more detail. In particular, we analyze the
above-mentioned potential energy shifts and their influence on the spectra,
\ for smaller internuclear distances than those taken in \cite{F2007}. We
also provide an alternative interpretation of the results encountered, based
on effective prefactors and single-atom saddle-point equations.

This paper is organized as follows. In Sec. \ref{transampl}, we briefly
recall the strong-field approximation HHG transition amplitudes. Thereby, we
consider the situation for which the structure of the molecule is either
incorporated in the prefactor (Sec. \ref{prefactor} ), or in the
semiclassical action (Sec. \ref{Smodified}). Subsequently (Sec. \ref{results}%
), we analyze the role of the different scenarios, involving one and two
centers, in the high-harmonic spectra, either solving the modified
saddle-point equations (Sec. \ref{orbits}), or mimicking the
quantum-interference between different sets of orbits by an adequate choice
of prefactors (Sec. \ref{singleatom}). Finally, in Sec. \ref{concl} we
outline the main conclusions of this work.

\section{Transition amplitudes}

\label{transampl}

\subsection{General expressions}

As a starting point, we will underline our main assumptions with regard to
the diatomic bound-state wave functions. We consider frozen nuclei, the
linear combination of atomic orbitals (LCAO) approximation, and homonuclear
molecules. Under these assumptions, the electronic bound-state wave function
reads\
\begin{equation}
\psi _{0}(\mathbf{r})=C_{\psi }(\phi _{0}(\mathbf{r}-\mathbf{R}/2)+\epsilon
\phi _{0}(\mathbf{r}+\mathbf{R}/2)),  \label{LCAO}
\end{equation}%
where $\epsilon =\pm 1,$ $C_{\psi }=1/\sqrt{2(1+\epsilon S(\mathbf{R})},$
with
\begin{equation}
S(\mathbf{R})=\int \left[ \phi _{0}(\mathbf{r}-\mathbf{R}/2)\right] ^{\ast
}\phi _{0}(\mathbf{r}+\mathbf{R}/2)d^{3}r
\end{equation}%
The positive and negative signs for $\epsilon $ denote symmetric and
antisymmetric orbitals, respectively. For simplicity, unless otherwise
stated we will consider parallel-aligned molecules.

The SFA transition amplitude for high-order harmonic generation reads, in
the specific formulation of Ref. \cite{hhgsfa} and in atomic units,
\begin{eqnarray}
M^{(\Omega )} &\hspace{-0.1cm}=\hspace*{-0.1cm}&i\int_{-\infty }^{\infty }%
\hspace*{-0.5cm}dt\int_{-\infty }^{t}~\hspace*{-0.5cm}dt^{\prime }\int
d^{3}kd_{\mathrm{rec}}^{\ast }(\mathbf{\tilde{k}}(t))d_{\mathrm{ion}}(%
\mathbf{\tilde{k}}(t^{\prime }))  \notag \\
&&\exp [iS(t,t^{\prime },\Omega ,\mathbf{k})]+c.c.,  \label{amplhhg}
\end{eqnarray}%
with the action
\begin{equation}
S(t,t^{\prime },\Omega ,\mathbf{k})=-\frac{1}{2}\int_{t^{\prime }}^{t}[%
\mathbf{k}+\mathbf{A}(\tau )]^{2}d\tau -I_{p}(t-t^{\prime })+\Omega t
\label{actionhhg}
\end{equation}%
and the prefactors $d_{\mathrm{rec}}(\mathbf{\tilde{k}}(t))=\left\langle
\mathbf{\tilde{k}}(t)\right\vert \mathbf{r}.\mathbf{e}_{x}\left\vert \psi
_{0}\right\rangle $ and $d_{\mathrm{ion}}(\mathbf{\tilde{k}}(t^{\prime
}))=\left\langle \mathbf{\tilde{k}}(t^{\prime })\right\vert H_{\mathrm{int}}%
\mathbf{(}t^{\prime }\mathbf{)}\left\vert \psi _{0}\right\rangle .$ Thereby $%
\mathbf{r}$, $\mathbf{e}_{x}$, $H_{\mathrm{int}}\mathbf{(}t^{\prime }\mathbf{%
),}$ $I_{p},$ and $\Omega $ give the dipole operator, the laser-polarization
vector, the interaction with the field, the ionization potential, and the
harmonic frequency, respectively. The explicit expressions for $\mathbf{%
\tilde{k}}(t)$ are gauge dependent, and will be provided below. Physically,
Eq. (\ref{amplhhg}) describes a process in which an electron, initially in a
field-free bound-state $\left\vert \psi _{0}\right\rangle $, is coupled to a
Volkov state $\left\vert \mathbf{\tilde{k}}(t^{\prime })\right\rangle $ by
the interaction $H_{\mathrm{int}}\mathbf{(}t^{\prime }\mathbf{)}$ of the
system with the field. Thereafter, it propagates in the continuum and is
driven back towards its parent ion, or molecule. At a time $t,$ it
recombines, emitting high-harmonic radiation of frequency $\Omega .$

The above-stated transition amplitude may be either solved
numerically, or employing saddle-point equations. In this work, we
employ the latter method and the specific uniform approximation
discussed in Ref. \cite{atiuni}. Explicitly, these equations are
given by the condition that the semiclassical action be stationary,
i.e., that $\partial _{t}S(t,t^{\prime
},\Omega ,\mathbf{k})=\partial _{t^{\prime }}S(t,t^{\prime },\Omega ,\mathbf{%
k})=0$ and $\partial _{\mathbf{k}}S(t,t^{\prime },\Omega ,\mathbf{k})=%
\mathbf{0.}$

For a single atom placed at the origin of the coordinate system, this leads
to
\begin{equation}
\left[ \mathbf{k}+\mathbf{A}(t^{\prime })\right] ^{2}=-2I_{p},
\label{saddle1}
\end{equation}%
\begin{equation}
\int_{t^{\prime }}^{t}d\tau \left[ \mathbf{k}+\mathbf{A}(\tau
)\right] =0, \label{saddle3}
\end{equation}%
and
\begin{equation}
2(\Omega -I_{p})=\left[ \mathbf{k}+\mathbf{A}(t)\right] ^{2}.
\label{saddle2}
\end{equation}%
Eq. (\ref{saddle1}) gives the conservation of energy at the instant $%
t^{\prime }$of ionization,\ and has no real solution. Indeed, the time $%
t^{\prime }$ will possess a non-vanishing imaginary part. This is due to the
fact that tunneling is a process which has no classical counterpart. In the
limit $I_{p}\rightarrow 0,$ corresponds to the physical situation of a
classical electron reaching the continuum with vanishing drift velocity. Eq.
(\ref{saddle3}) expresses the fact that the electron propagates in the
continuum from $t^{\prime }$ to $t,$ when it returns to the site of its
release. Eq. (\ref{saddle2}) yields the conservation of energy at the
recombination instant $t,$ when the kinetic energy of the returning electron
is converted into high-order harmonic radiation.

One should note that the transition amplitude (\ref{amplhhg}) is gauge
dependent \cite{FKS96,PRACL2006}. Firstly, the interaction Hamiltonians $H_{%
\mathrm{int}}(t^{\prime })$, which are present in $d_{\mathrm{ion}}(\mathbf{%
\tilde{k}}(t^{\prime }))$, are different in the length and velocity gauges.
Furthermore, in both velocity- and length-gauge formulations, field-free
bound states are taken, which are not gauge equivalent. Therefore, different
gauge choices will yield different interference patterns \cite%
{PRACL2006,DM2006,Madsen,Usachenko,F2007,SSY2007}. This problem has been
overcome to a large extent by considering field-dressed bound states, as a
dressed state in the length gauge is gauge-equivalent to a field-free bound
state in the velocity gauge, and vice-versa (for details see \cite%
{dressedSFA,F2007,DM2006}).

\subsection{Double-slit interference condition}

\label{prefactor}

The matrix element $d_{\mathrm{rec}}(\mathbf{\tilde{k}})=\left\langle
\mathbf{\tilde{k}}\right\vert \mathbf{r}\cdot \mathbf{e}_{x}\left\vert \psi
_{0}\right\rangle $ is then given by
\begin{equation}
d_{\mathrm{rec}}^{(b)}(\mathbf{\tilde{k}})=\frac{2iC_{\psi }}{(2\pi )^{3/2}}%
\left[ -\cos (\vartheta )\partial _{p_{x}}\phi (\mathbf{\tilde{k}})+\frac{%
R_{x}}{2}\sin (\vartheta )\phi (\mathbf{\tilde{k}})\right] ,  \label{prefb}
\end{equation}%
for bonding molecular orbitals (i.e., $\epsilon >0),$ or
\begin{equation}
d_{\mathrm{rec}}^{(a)}(\mathbf{\tilde{k}})=\frac{2C_{\psi }}{(2\pi )^{3/2}}%
\left[ \sin (\vartheta )\partial _{p_{x}}\phi (\mathbf{\tilde{k}})-\frac{%
R_{x}}{2}\cos (\vartheta )\phi (\mathbf{\tilde{k}})\right] ,  \label{prefa}
\end{equation}%
in the antibonding case (i.e., $\epsilon <0),$ with $\vartheta =\mathbf{%
\tilde{k}}\cdot \mathbf{R}/2.$ In the above-stated equations, $R_{x}$
denotes the projection of the internuclear distance along the direction of
the laser-field polarization.

In Eqs. (\ref{prefb}) and (\ref{prefa}), the terms with a purely
trigonometric dependence on the internuclear distance yield the double-slit
condition in \cite{doubleslit}. The maxima and minima in the spectra which
are caused by this condition are expected to occur for
\begin{equation}
\mathbf{\tilde{k}}\cdot \mathbf{R}=2n\pi \text{ and }\mathbf{\tilde{k}}\cdot
\mathbf{R}=(2n+1)\pi ,  \label{maxmin}
\end{equation}%
respectively, for bonding molecular orbitals (i.e., $\epsilon >0).$ For
antibonding orbitals, the maxima occur for the odd multiples of $\pi $ and
the minima for the even multiples. In the length and velocity gauges $\mathbf{%
\tilde{k}}(\tau)=\mathbf{k}+\mathbf{A}(\tau)$ and
$\mathbf{\tilde{k}}(\tau)=\mathbf{k} $, where $\tau=t,t^{\prime}$,
respectively.

The remaining terms grow linearly with $R_{x}$, and are an artifact of the
strong-field approximation, due to the fact that the continuum states and
the bound states are not orthogonal in the context of the strong-field
approximation \cite{JMOCL2006,F2007,SSY2007}. For that reason, they will be
neglected here (for rigorous justifications see \cite{DM2006,SSY2007}).

In the length gauge, $d_{\mathrm{rec}}(\mathbf{\tilde{k}}(t))=d_{\mathrm{ion}%
}(\mathbf{\tilde{k}}(t^{\prime }))$, with $\ \mathbf{\tilde{k}}(t)=\mathbf{k}%
+\mathbf{A}(t),$ while in the velocity gauge,
\begin{equation}
d_{\mathrm{ion}}^{(b)}(\mathbf{\tilde{k}})=\frac{C_{\psi }[\mathbf{k}+%
\mathbf{A}(t^{\prime })]^{2}}{(2\pi )^{3/2}}\cos (\vartheta )\phi (\mathbf{%
\tilde{k}}),
\end{equation}%
or%
\begin{equation}
d_{\mathrm{ion}}^{(a)}(\mathbf{\tilde{k}})=-i\frac{C_{\psi }[\mathbf{k}+%
\mathbf{A}(t^{\prime })]^{2}}{(2\pi )^{3/2}}\sin (\vartheta )\phi (\mathbf{%
\tilde{k}}),
\end{equation}%
with $\mathbf{\tilde{k}}(t)=\mathbf{k,}$ for bonding and antibonding
molecular orbitals, respectively. The simplest and most widely adopted \cite%
{MBBF00,Madsen,Usachenko,KB2005,DM2006,JMOCL2006} procedure is to employ the
prefactors $d_{\mathrm{ion}}(\mathbf{\tilde{k}})$ $\ $and $d_{\mathrm{rec}}(%
\mathbf{\tilde{k}})$ and the single-atom saddle-point equations (\ref%
{saddle1})-(\ref{saddle2}). In this case, we consider the origin, from which
the electron leaves and returns, as the geometric center of the molecule.

\subsection{Modified saddle-point equations}

\label{Smodified}

The prefactors $d_{\mathrm{ion}}^{(b)}(\mathbf{\tilde{k}})$ and $d_{\mathrm{%
rec}}^{(b)}(\mathbf{\tilde{k}})$ will now be exponentialized and
incorporated in the action (for details, see \cite{PRACL2006,F2007}). For
the recombination matrix element, we take the expression
\begin{equation}
d_{\mathrm{rec}}^{(b)}(\mathbf{\tilde{k}})=-\frac{2iC_{\psi }}{(2\pi )^{3/2}}%
\left[ \cos \left( \mathbf{\tilde{k}}\cdot \frac{\mathbf{R}}{2}\right)
\partial _{\tilde{p}_{x}}\phi (\mathbf{\tilde{k}})\right] ,
\label{modifieddip}
\end{equation}%
for which the spurious term is $R_{x}$ is absent. In the expression for the
antibonding case, the cosine term in (\ref{modifieddip}) should be replaced
by $\sin (\mathbf{\tilde{k}\cdot R}/2)$. Without loss of generality, the
same procedure can also be applied to more complex orbitals.

This leads to the sum
\begin{equation}
M=\sum_{j=1}^{2}\sum_{\nu =1}^{2}M_{j\nu } \label{sumampl}
\end{equation}%
of the transition amplitudes
\begin{eqnarray}
M_{j\nu } &=&\frac{C_{\psi }}{(2\pi )^{3/2}}\int_{0}^{t}dt^{%
\prime }\int dt\int d^{3}p\eta (\mathbf{k},t,t^{\prime })  \notag \\
&&\times \exp [iS_{j\nu }(\mathbf{k},\Omega ,t,t^{\prime })],\
\label{amplitudes}
\end{eqnarray}%
with $\eta (\mathbf{k},t,t^{\prime })=\left[ \partial _{\tilde{p}_{x}}\phi (%
\mathbf{\tilde{k}}(t))\right] ^{\ast }\partial _{\tilde{p}_{x}}\phi (\mathbf{%
\tilde{k}(}t^{\prime })).$ The terms $S_{j\nu }(\mathbf{k},\Omega
,t,t^{\prime })$ correspond to a modified action, which incorporates the
structure of the molecule. Explicitly, they read
\begin{equation}
S_{j\nu }(\mathbf{k},\Omega ,t,t^{\prime })=S(\mathbf{k},\Omega ,t,t^{\prime
})+(-1)^{\nu +1}\xi (R,t,t^{\prime })  \label{ssame}
\end{equation}%
where $\xi (R,t,t^{\prime })=[\mathbf{\tilde{k}}(t)\mathbf{-}(-1)^{\nu +j}%
\mathbf{\tilde{k}}(t^{\prime })]\cdot \mathbf{R}/2$.

We will now compute the amplitudes $M_{j\nu }$ employing saddle-point
methods. For this purpose, we will seek values for $t,t^{\prime }$ and $%
\mathbf{k}$ which satisfy the conditions $\partial _{\mathbf{k}}S_{j\nu }(%
\mathbf{k},\Omega ,t,t^{\prime })=\mathbf{0},\ \partial _{t}S_{j\nu }(%
\mathbf{k},\Omega ,t,t^{\prime })=0$ and $\partial _{t^{\prime }}S_{j\nu }(%
\mathbf{k},\Omega ,t,t^{\prime })=0$. This leads to the saddle-point
equations
\begin{equation}
\frac{\lbrack \mathbf{k}+\mathbf{A}(t^{\prime })]^{2}}{2}=-I_{p}+(-1)^{2\nu
+j+1}\partial _{t^{\prime }}\mathbf{\tilde{k}}(t^{\prime })\cdot \mathbf{R}%
/2,  \label{tunnel}
\end{equation}%
\begin{equation}
\int_{t^{\prime }}^{t}[\mathbf{k}+\mathbf{A}(s)]ds+(-1)^{\nu +1}\partial _{%
\mathbf{k}}\zeta =0,  \label{returndiff}
\end{equation}%
with $\zeta =\left[ \mathbf{\tilde{k}}(t)-(-1)^{j+\nu }\mathbf{\tilde{k}}%
(t^{\prime })\right] \cdot \mathbf{R}/2$ and%
\begin{equation}
\frac{\lbrack \mathbf{k}+\mathbf{A}(t)]^{2}}{2}=\Omega -I_{p}+(-1)^{\nu
}\partial _{t}\mathbf{\tilde{k}}(t)\cdot \mathbf{R}/2.  \label{rec}
\end{equation}%
Eq. (\ref{tunnel}) corresponds to the tunnel ionization process,
saddle-point equation (\ref{returndiff}) gives the condition that the
electron returns to its parent molecule and Eq. (\ref{rec}) expresses the
conservation of energy at the instant of recombination, in which the kinetic
energy of the electron is converted into high-order harmonic radiation. The
above-stated saddle-point equations depend on the gauge, on the center $%
C_{j} $ from which the electron was freed and on the center $C_{\nu }$ with
which it recombines. Below we will have a closer look at specific cases. We
will start by analyzing Eqs. (\ref{tunnel}) and (\ref{rec}), which,
physically, correspond to the ionization and recombination process,
respectively.

If the length gauge is chosen, both equations are explicitly written as
\begin{equation}
\frac{\lbrack \mathbf{k}+\mathbf{A}(t^{\prime })]^{2}}{2}=-I_{p}+(-1)^{2\nu
+j}\mathbf{E}(t^{\prime })\cdot \mathbf{R}/2,
\end{equation}%
and%
\begin{equation}
\frac{\lbrack \mathbf{k}+\mathbf{A}(t)]^{2}}{2}=\Omega -I_{p}+(-1)^{\nu +1}%
\mathbf{E}(t)\cdot \mathbf{R}/2,
\end{equation}%
respectively. For this specific formulation, there exist potential-energy
shifts on the right-hand side, which depend on the external laser field $%
\mathbf{E}(\tau )(\tau =t,t^{\prime })$ and on the internuclear distance $%
\mathbf{R.}$ At the ionization or recombination times, depending on the
center, they increase, or sink the potential-energy barrier through which
the electron must tunnel, or the energy of the state with which it will
recombine. In the specific case discussed here, there is a decrease in the
barrier at $C_{2}$ and an increase at $C_{1}.$ Their meaning and existence
altogether has raised considerable debate in the literature \cite%
{PRACL2006,DM2006,SSY2007,BCCM2007}.

In the velocity gauge, the saddle-point equations (\ref{tunnel}) and (\ref%
{rec}) read
\begin{equation}
\frac{\lbrack \mathbf{k}+\mathbf{A}(t^{\prime })]^{2}}{2}=-I_{p},
\label{tunnelv}
\end{equation}%
and%
\begin{equation}
\frac{\lbrack \mathbf{k}+\mathbf{A}(t)]^{2}}{2}=\Omega -I_{p}.  \label{recv}
\end{equation}%
These equations do not exhibit the above-mentioned potential-energy shifts,
and resemble the saddle-point equations obtained for a single atom \cite%
{hhgsfa}. Furthermore, if the limit $I_{p}\rightarrow 0$ is taken, Eq.(\ref%
{tunnelv}) describes a classical particle reaching the continuum with
vanishing drift momentum. In contrast, in the length gauge neither the
classical limit nor the single-atom equations are obtained.
\begin{figure}[tbp]
\begin{center}
\includegraphics[width=9cm]{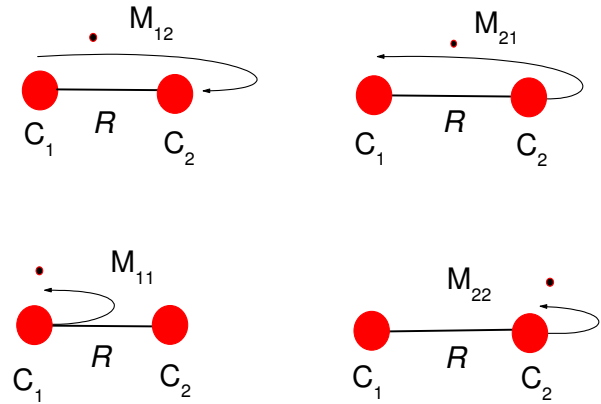}
\end{center}
\caption{Schematic representation of the four possible recombination or
rescattering scenarios described by Eq. (\protect\ref{sumampl}). The centers
$C_{1}$ and $C_{2}$ in the molecule, as well as the transition amplitudes $%
M_{j\protect\nu }$ ($j,\protect\nu =1,2$) are indicated in the figure.}
\label{contour}
\end{figure}

We will now discuss the saddle-point equation (\ref{returndiff}) which gives
the return condition. For both length and velocity gauges, one may
distinguish two main scenarios: either the electron leaves and returns to
the same center, i.e., $\nu =j$, or the electron is freed at a center $C_{j}$
and recombines with the other center $C_{\nu }$, $j\neq \nu ,$ in the
molecule. In the former and latter case, the return condition reads%
\begin{equation}
\int_{t^{\prime }}^{t}[\mathbf{k}+\mathbf{A}(s)]ds=0,  \label{return1C}
\end{equation}%
or
\begin{equation}
\int_{t^{\prime }}^{t}[\mathbf{k}+\mathbf{A}(s)]ds+(-1)^{\nu +1}\mathbf{R}=0.
\label{return2C}
\end{equation}%
In Eq. (\ref{return2C}), the index $\nu =2$ corresponds to the transition
amplitudes $M_{12}$ (center $C_{1}$ to center $C_{2})$ and $M_{21}$ (center $%
C_{2}$ to center $C_{1})$, respectively. For clarity, the scenarios
described above are summarized in Fig. 1.

\section{Quantum interference and different recombination scenarios}

\label{results} In the following we will discuss high-order harmonic
spectra. For simplicity, we will consider that the electrons involved are
initially bound in $1s$ states. This gives
\begin{equation}
\phi (\mathbf{\tilde{k}})\sim \frac{1}{[\mathbf{\tilde{k}}^{2}+2I_{p}]^{2}}
\label{dip1s}
\end{equation}%
in the high-order harmonic prefactors $d_{\mathrm{ion}}^{(b)}(\mathbf{\tilde{%
k}})$ and $d_{\mathrm{rec}}^{(b)}(\mathbf{\tilde{k}})$.

\begin{figure}[tbp]
\begin{center}
\noindent \includegraphics[width=9cm]{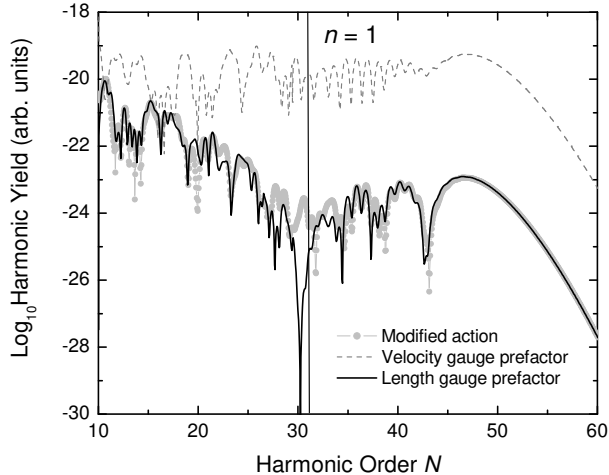}
\end{center}
\caption{Spectra computed employing the single-atom orbits and two center
prefactors, for the length and velocity gauges, compared to the length-gauge
spectrum obtained employing modified saddle-point equations. We consider
here the modified length form ({\protect\ref{modifieddip}}) of the dipole
operator, which excludes the term with a linear dependence on $R_{x}$. The
atomic system was approximated by the linear combination of $1s$ atomic
orbitals with $I_{p}=0.57$ a.u.. The internuclear distance and the alignment
angle are $R=2.068$ a.u., and $\protect\theta =0,$ respectively. The driving
field intensity and frequency are given by $I=3\times 10^{14}\mathrm{W/cm^{2}%
}$, and $\protect\omega =0.057$ a.u., respectively. The interference minimum
at $n=1$ is indicated by the vertical line in the figure. The difference in
the orders of magnitude between the velocity and length gauge spectra is due
to the different prefactors $d_{\mathrm{ion}}(\tilde{\mathbf{k}})$.}
\label{interfe1}
\end{figure}

In Fig. \ref{interfe1}, we will commence by displaying the overall
contributions, computed using the prefactors $d_{\mathrm{ion}}^{(b)}(\mathbf{%
\tilde{k}})$ and $d_{\mathrm{rec}}^{(b)}(\mathbf{\tilde{k}})$ and
single-atom saddle point equations, instead of the modified saddle-point
equations (\ref{tunnel})-(\ref{rec}), for the length and velocity gauges.
For comparison, the also present the contribution from all transition
amplitudes $M_{j\nu }^{(\Omega )}$. In the present computations, we
considered up to five pairs of orbits starting at the first half-cycle of
the field, i.e., $0\leq t^{\prime }\leq T/2.$

In the length gauge, the interference condition predicts interference
extrema at $\Omega =I_{p}+n^{2}\pi ^{2}/(2R^{2})$. For the parameters in the
figure, this yields a minimum near $\Omega =31\omega $, for $n=1.$ Even
though this minimum is shallower if modified saddle-point equations are
taken, it can be easily identified.

In contrast, in the velocity gauge, the above-mentioned interference
patterns are absent. This is due to the fact that the interference condition
changes. The maxima and minima re now given by (\ref{maxmin}), with $\mathbf{%
k}$ instead of $\mathbf{\tilde{k}}=\mathbf{k}+\mathbf{A}(t).$ This will lead
to interference extrema at harmonic frequency $\Omega =I_{p}+\left[ n^{2}\pi
^{2}/R^{2}+2n\pi A(t)/R+A^{2}(t)\right] /2.$ Roughly, if we assume that the
vector potential at the electron return time is $A(t)\simeq 2\sqrt{U_{p}},$
this will correspond to $\Omega \sim 97\omega .$ This frequency lies far
beyond the cutoff ( $\Omega \sim 47\omega )$, so that there will be a
breakdown in the interference patterns \cite{F2007,SSY2007}. For this
reason, in the following figures we will consider only the length-gauge
situation.

\subsection{Modified saddle-point equations}

\label{orbits}

\begin{figure}[tbp]
\begin{center}
\noindent \includegraphics[width=9cm]{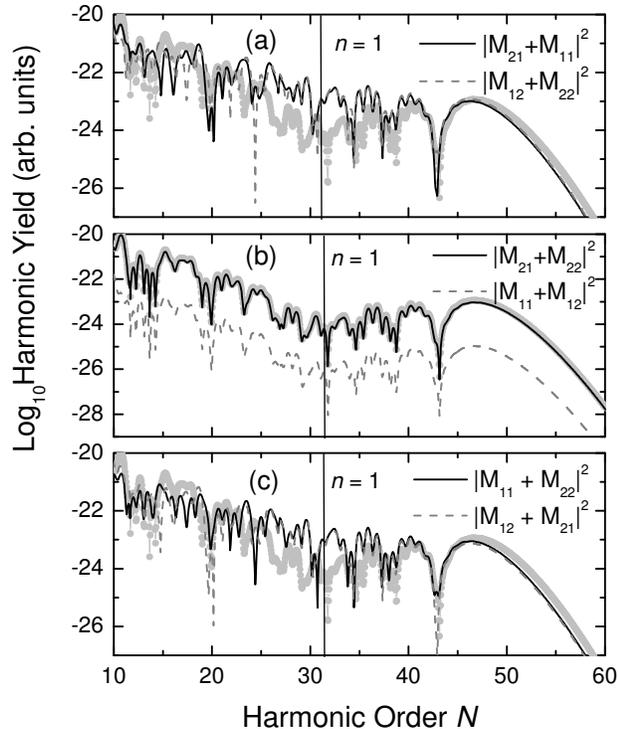}
\end{center}
\caption{Contributions to the high-harmonic yield from the quantum
interference between different types of orbits, for internuclear distance $%
R=2.068$ a.u. The remaining parameters are the same as in Fig. 2. Panel (a):
Orbits involving similar scattering scenarios, i. e., $|M_{11}+M_{22}|^{2}$,
and $|M_{12}+M_{21}|^{2}.$ Panel (b): Orbits \emph{starting} at the same
center, i.e., $|M_{11}+M_{12}|^{2}$ and $|M_{21}+M_{22}|^{2}.$ Panel (b):
Orbits \emph{ending} at the same center, i.e., $|M_{11}+M_{21}|^{2}$ and $%
|M_{12}+M_{22}|^{2}.$ For comparison, the full contributions $%
|M_{21}+M_{22}+M_{11}+M_{12}|^{2}$ are displayed as the light gray circles
in the picture. The interference minimum at $n=1$ is indicated as the
vertical line in the figure.}
\label{orbits1}
\end{figure}

Subsequently, in Fig. \ref{orbits1}, we present the contributions from the
different recombination scenarios. In panel (a), the contributions from the
topologically similar scenarios, involving only one or two centers, are
depicted. We observe that the interference minimum mentioned in Fig. 1 is
absent for both types of contributions. \ At first sight, this seems to
contradict the double-slit picture. In fact, for both $|M_{12}+M_{21}|^{2}$
and $|M_{11}+M_{22}|^{2},$ high-order harmonic emission at spatially
separated centers takes place. Therefore, one would expect well-defined
interference patterns to be present. One should note, however, that the
potential energy shifts $\pm \mathbf{E}(t^{\prime })\cdot \mathbf{R}/2$ sink
the potential barrier for the orbits starting at $C_{2}$ and increase the
potential barrier for those starting at $C_{1}.$ Thus, the latter
contributions are strongly suppressed and do not contribute significantly to
the two-center interference.

This is in agreement with panel (b), in which the contributions from the
processes $|M_{jj}+M_{j\nu }|^{2}(j,\nu =1,2$ and $\nu \neq j)$ starting
from the same center and ending at different centers are depicted. Therein,
the contributions of the processes starting at $C_{2}$ are roughly two
orders of magnitude larger than those starting at $C_{1}.$ This is due to
the fact that the barrier through which the electron must tunnel in order to
reach the continuum is much wider for the latter center. Furthermore, the
two-center interference minimum near $\Omega =31\omega $ is present. This is
expected, as the contributions from the centers $C_{1}$ and $C_{2}$ exhibit
the same order of magnitude for both types of orbits.

Finally, in panel (c) we display the contributions $|M_{jj}+M_{\nu j}|^{2}$
from the processes starting at different centers and ending at the same
center. In this case, the interference minimum is absent. This was expected
for two reasons. First, for these orbits, there is no high-order harmonic
emission taking place at spatially separated centers. Second, even if this
were the case, the contributions from the orbits starting at $C_{2}$ are
much stronger than those starting at $C_{1}$.

\begin{figure}[tbp]
\begin{center}
\noindent \includegraphics[width=9cm]{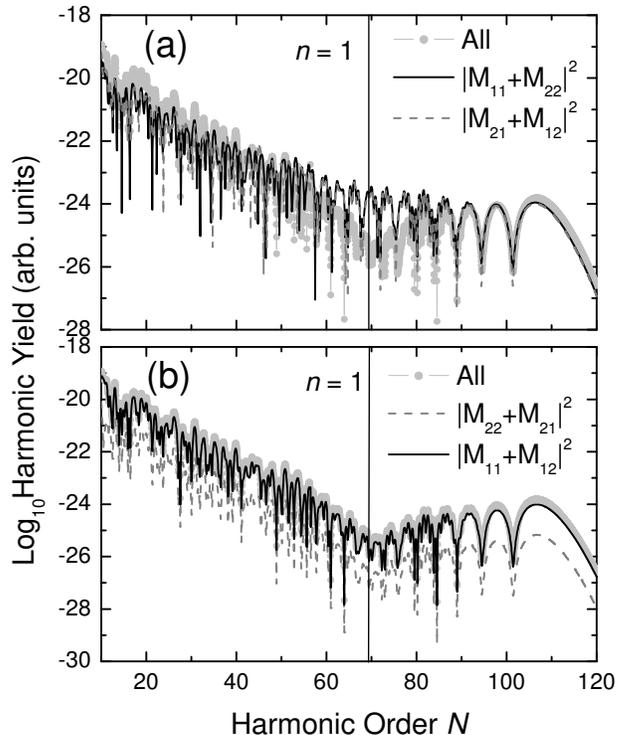}
\end{center}
\caption{Contributions from different types of orbits to the high-harmonic
yield, for internuclear distance $R=1.2$ a.u. and intensity $I=8\times
10^{14}\mathrm{W/cm}^{2}.$ The remaining parameters are the same as in the
previous figures. Panel (a) gives the contributions from the topologically
similar scattering scenarios, i. e., $|M_{11}+M_{22}|^{2}$, and $%
|M_{12}+M_{21}|^{2},$\ and panel (b) of the orbits starting at the same
center, i.e., $|M_{11}+M_{12}|^{2}$ and $|M_{21}+M_{22}|^{2}.$}
\label{smallR1}
\end{figure}

Since the potential energy shifts $\pm \mathbf{E}(t^{\prime })\cdot \mathbf{R%
}/2$ depend on the internuclear distance, it is legitimate to ask the
question of whether, for small internuclear distances, a minimum is present
in the contributions from the topologically similar scenarios.\ In Fig. \ref%
{smallR1}, we considered such a situation. From the interference condition,
we expect a minimum near $\Omega =69\omega $. This minimum is present for
the overall contributions, and also for the processes $|M_{jj}+M_{j\nu
}|^{2}(j,\nu =1,2$ and $\nu \neq j)$ starting from the same center and
ending at different centers [Fig. \ref{smallR1}.(a)]. It is however absent
for the interference of topologically similar processes [Fig. \ref{smallR1}%
.(b)]. This is due to fact that, even for this small internuclear distance,
the orbits starting from $C_{2}$ lead to larger contributions than those
starting from $C_{1}.$ Indeed, a closer look at Fig. \ref{smallR1}.(a) shows
that the contributions $|M_{11}+M_{12}|^{2}$ are roughly one order of
magnitude smaller than $|M_{22}+M_{21}|^{2}$.

Possibly, in order to obtain well-defined maxima and minima for the
contributions of topologically similar scenarios, it would be necessary to
reduce the internuclear distance even more. In this case, however, none of
the assumptions adopted in this paper, such as the LCAO approximation, hold.
In this context, it is worth noticing that the parameters adopted in Fig. %
\ref{smallR1} are also somewhat unrealistic, as far as this specific
approximation is concerned. If, however, an alternative ionization pathway
is provided, so that the electron may reach the continuum without the need
of overcoming the potential-energy barriers, the contributions from the
topologically similar scenarios may lead to well-defined patterns. Indeed,
in previous work, we employed an additional attosecond-pulse train in order
to release the electron in the continuum, and obtained an interference
minimum in this case \cite{F2007}. We were, however, changing the physics of
the problem by providing a different ionization mechanism. In the following,
we will investigate the issue of the potential-energy shifts for this set of
parameters, employing an alternative method.

\subsection{Modified prefactors}

\label{singleatom}

On the other hand, the transition amplitudes $M_{j\nu }$ may also be grouped
in such a way as to obtain effective prefactors. Such prefactors may then be
related to the quantum interference of specific types of orbits. Hence, one
may mimic the influence of the above-stated scenarios even if the
single-atom saddle-point equations (\ref{saddle1})-(\ref{saddle2}) are taken
into account. For the symmetric combination of atomic orbitals considered
here, there would be four different sets of prefactors, which are explicitly
given by%
\begin{eqnarray}
d_{\mathrm{ion}}^{(j\nu )}(\mathbf{k},t,t^{\prime }) &=&2\exp [(-1)^{j}i%
\mathbf{\tilde{k}}(t^{\prime })\cdot \mathbf{R}/2]  \label{samestart} \\
&&\times \cos [\mathbf{\tilde{k}}(t)\cdot \mathbf{R}/2]\eta (\mathbf{k}%
,t,t^{\prime }),  \notag
\end{eqnarray}%
\begin{eqnarray}
d_{\mathrm{end}}^{(j\nu )}(\mathbf{k},t,t^{\prime }) &=&2\exp [(-1)^{j}i[%
\mathbf{\tilde{k}}(t)\mathbf{+A}(t^{\prime })]\cdot \mathbf{R}/2]
\label{sameend} \\
&&\times \cos [\mathbf{k}\cdot \mathbf{R}/2]\eta (\mathbf{k},t,t^{\prime }),
\notag
\end{eqnarray}%
\begin{equation}
d_{\mathrm{same}}(\mathbf{k},t,t^{\prime })=2\cos [\mathbf{[A}(t)-\mathbf{A}%
(t^{\prime })]\cdot \mathbf{R}/2]\eta (\mathbf{k},t,t^{\prime }),
\end{equation}%
and

\begin{eqnarray}
d_{\mathrm{diff}}(\mathbf{k},t,t^{\prime }) &=&2\cos [\mathbf{p}\cdot
\mathbf{R}+[\mathbf{A}(t)+\mathbf{A}(t^{\prime })]\cdot \mathbf{R}/2] \\
&&\times \eta (\mathbf{k},t,t^{\prime }).  \notag
\end{eqnarray}

\begin{figure}[tbp]
\begin{center}
\noindent \includegraphics[width=9cm]{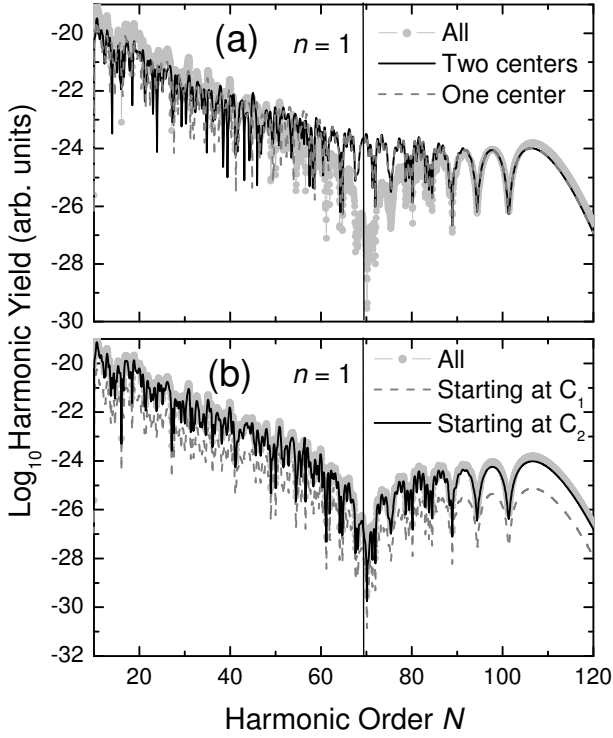}
\end{center}
\caption{Contributions from different types of orbits to the high-harmonic
yield, for the same parameters as in Fig. 4. Panel (a) gives the
contributions from the topologically similar scattering scenarios, i. e., $%
|M_{11}+M_{22}|^{2}$, and $|M_{12}+M_{21}|^{2},$\ and panel (b) of the
orbits starting at the same center, i.e., $|M_{11}+M_{12}|^{2}$ and $%
|M_{21}+M_{22}|^{2}.$ All results in this figure have been computed
mimicking the above-stated processes by employing the modified prefactors
(27)-(30) and single-atom saddle-point equations.}
\label{smallR2}
\end{figure}
The prefactor $d_{\mathrm{ion}}^{(j\nu )}$ corresponds to the transition
amplitudes $M_{jj}+M_{\nu j}$ in which the electron starts at the same
center and recombines with different centers in the molecule. The prefactor $%
d_{\mathrm{end}}^{(j\nu )}$ is related to the transition amplitudes $M_{\nu
j}+M_{jj}$ in which the electron starts at different centers, but ends at
the same center $C_{j}$. Finally, $d_{\mathrm{same}}$ and $d_{\mathrm{diff}}$
corresponds to the topologically similar processes, in which only one, or
two center scenarios, respectively, are involved. Interestingly, only the
prefactors $d_{\mathrm{ion}}^{(j\nu )}$ lead to the same interference
conditions as the overall double-slit prefactor (\ref{prefb}).

Furthermore, one should note that, if all parameters involved were real, for
the first two prefactors there would be the symmetry $|d_{\mathrm{ion}%
}^{(j\nu )}(\mathbf{k},t,t^{\prime })|^{2}=|d_{\mathrm{ion}}^{(\nu j)}(%
\mathbf{k},t,t^{\prime })|^{2}$ and $|d_{\mathrm{end}}^{(j\nu )}(\mathbf{k}%
,t,t^{\prime })|^{2}=|d_{\mathrm{end}}^{(\nu j)}(\mathbf{k},t,t^{\prime
})|^{2}.$ This would lead to the same transition probabilities, as one
transition amplitude is the complex conjugate of the other. This is,
however, not the case, and can be seen by inspecting Eq. (\ref{samestart}).
Specifically in the length gauge, $\mathbf{\tilde{k}}(t^{\prime })\mathbf{%
=k+A}(t^{\prime })$. Depending on the center, this will lead to
exponentially decreasing or increasing factors $\exp [\mp \mathrm{Im}%
[k+A(t^{\prime })]R]$ in the transition probability $|M_{jj}+M_{\nu j}|^{2}.$
Clearly, this procedure is less rigorous than that adopted in the previous
section, as we are not considering the influence of the potential-energy
shifts in the imaginary part of $t^{\prime }.$

In Fig. \ref{smallR2}, we display the results obtained following the
above-stated procedure, for the same parameters as in Fig.\ref{smallR1}.
Once more, we see that the contributions of topologically similar processes,
involving either one or two centers, do not lead to a well-defined
interference minimum (Fig. \ref{smallR2}.(a)). Additionally, the quantum
interference of the two different kinds of processes starting from the same
center $C_{j}$ leads to a well-defined minimum at the expected frequency $%
\Omega =69\omega $. Furthermore, the contributions from the orbits starting
at $C_{2}$ are also roughly one order of magnitude smaller. The main
difference between the two approaches is that the interference minimum is
much deeper if modified prefactors are taken, as compared with the results
obtained with modified saddle-point equations. This discrepancy is present
throughout, and has also been observed in Ref. \cite{F2007}.

\section{Conclusions}

\label{concl} The results presented in this work indicate that the
double-slit interference maxima and minima in the high-order harmonic
spectra, which are attributed to HHG at spatially separated centers, are
mainly due to the quantum interference between the processes $%
|M_{jj}+M_{j\nu }|^{2}$, $(j=1,2)$ in which the electron is released in the
continuum at a center $C_{j}$ in the molecule, and, subsequently, recombine
either at the same center or at a different center $C_{\nu }$. This can be
seen either by employing modified saddle-point equations, in which the
one-or two center scenarios are incorporated in the action, or by utilizing
modified prefactors in which only the above-stated processes are included.
In particular, when using the latter method, the transition amplitudes
related to both processes can be grouped in such a way that the
corresponding prefactor $d_{\mathrm{ion}}^{(j\nu )}(\mathbf{k},t,t^{\prime })
$ exhibits the same interference conditions as those in the overall
prefactor (\ref{prefb}). This is in agreement with the results obtained in
\cite{F2007}.

These results are not obvious, as there are other processes which lead to
high-order harmonic emission at different centers in the molecule. They do
not lead, however, to the double-slit interference patterns. This is due to
the fact that, in the present framework, there exist potential-energy shifts
that, depending on the center, sink or increase the barrier through which
the electron must initially tunnel. Therefore, they strongly suppress the
contributions to the spectra from one of the centers in the molecule. This
will lead to an absence of the two-center interference patterns for
processes starting at different centers. We have verified that this
suppression occurs even for small internuclear separations.

Such potential-energy shifts, however, are only present in the length-gauge
strong-field approximation and have raised a great deal of controversy \cite%
{PRACL2006,F2007,BCCM2007,SSY2007}. In fact, it is not even clear whether
they are not an artifact of the SFA. On the other hand, even if  single-atom
saddle-point equations are taken, we found a suppression in the yield for
one of the centers of the molecule. This in principle counterintuitive
result is related to the fact that the electron start time $t^{\prime }$ has
a non-vanishing imaginary part, which suppresses or enhances the yield
through the corresponding prefactors.

\acknowledgments This work has been financed by the UK EPSRC (Advanced
Fellowship, Grant no. EP/D07309X/1).

\end{document}